\documentclass[]{spie}  

\usepackage{amsmath,amsfonts,amssymb}
\usepackage{graphicx}
\usepackage{hyperref}
\hypersetup{colorlinks, allcolors=blue}
\usepackage{adjustbox}
\usepackage{caption}
\usepackage{subcaption}
\usepackage{multirow}
\usepackage{upgreek}
\usepackage[dvipsnames,table]{xcolor}
\usepackage{placeins}

\definecolor{matblue}{rgb}{0 0.447 0.741}
\definecolor{matred}{rgb}{0.85 0.3250 0.098}

\title{Improved Polarization Calibration \\ of the BICEP3 CMB Polarimeter at the South Pole}

\author[a]{J.~Cornelison}
\author[a]{C.~Vergès}

\author[b]{P.A.R. Ade}
\author[c]{Z. Ahmed}
\author[d]{M. Amiri}
\author[a]{D. Barkats}
\author[e]{R. Basu Thakur}
\author[c, f]{D. Beck}
\author[g]{C.A. Bischoff}
\author[e, h]{J.J. Bock}
\author[i]{V. Buza} 
\author[j]{J.R. Cheshire IV}
\author[k]{J. Connors}
\author[l]{M. Crumrine}
\author[e, f, c]{A.J. Cukierman}
\author[k]{E.V. Denison}
\author[a]{M.I. Dierickx}
\author[m]{L. Duband}
\author[a]{M. Eiben}
\author[d]{S. Fatigoni}
\author[n, o]{J.P. Filippini}
\author[g]{C. Giannakopoulos}
\author[f]{N. Goeckner-Wald}
\author[a]{D.C. Goldfinger}
\author[f]{J. Grayson}
\author[a]{P.K. Grimes}
\author[l]{G. Hall}
\author[f]{G. Halal}
\author[d]{M. Halpern}
\author[g]{E. Hand}
\author[a]{S.A. Harrison}
\author[c]{S. Henderson}
\author[e, h]{S.R. Hildebrandt}
\author[k]{G.C. Hilton}
\author[k]{J. Hubmayr}
\author[e]{H. Hui}
\author[f, c, k]{K.D. Irwin}
\author[f, e]{J. Kang}
\author[a, i]{K.S. Karkare}
\author[e]{S. Kefeli}
\author[a, p]{J.M. Kovac}
\author[f, c]{C.L. Kuo}
\author[l]{K. Lau}
\author[i]{E.M. Leitch}
\author[n]{A. Lennox}
\author[f]{T. Liu}
\author[a]{K. Look}
\author[h]{K.G. Megerian}
\author[e]{L. Minutolo}
\author[e]{L. Moncelsi}
\author[f]{Y. Nakato}
\author[q]{T. Namikawa}
\author[h]{H.T. Nguyen}
\author[e, h]{R. O'Brient}
\author[g]{S. Palladino}
\author[a]{M.A. Petroff}
\author[m]{T. Prouve}
\author[l, j]{C. Pryke}
\author[a ,r]{B. Racine}
\author[k]{C.D. Reinsema}
\author[f]{M. Salatino}
\author[e]{A. Schillaci}
\author[a]{B.L. Schmitt}
\author[j]{B. Singari}
\author[e]{A. Soliman}
\author[a, p]{T. St. Germaine}
\author[e]{B. Steinbach}
\author[b]{R.V. Sudiwala}
\author[f, c]{K.L. Thompson}
\author[a]{C. Tsai}
\author[b]{C. Tucker}
\author[h]{A.D. Turner}
\author[g, n]{C. Umilt\`{a}}
\author[s, i]{A.G. Vieregg}
\author[e]{A. Wandui}
\author[h]{A.C. Weber}
\author[d]{D.V. Wiebe}
\author[l]{J. Willmert}
\author[c]{W.L.K. Wu}
\author[f]{H. Yang}
\author[f, c]{K.W. Yoon}
\author[f, c]{E. Young}
\author[f]{C. Yu}
\author[a]{L. Zeng}
\author[e]{C. Zhang}
\author[e]{S. Zhang}

\affil[a]{Center for Astrophysics $\vert$ Harvard \& Smithsonian, Cambridge, MA 02138, U.S.A}
\affil[b]{School of Physics and Astronomy, Cardiff University, Cardiff, CF24 3AA, United Kingdom}
\affil[c]{Kavli Institute for Particle Astrophysics and Cosmology, SLAC National Accelerator Laboratory, Menlo Park, California 94025, USA}
\affil[d]{Department of Physics and Astronomy, University of British Columbia, Vancouver, British Columbia V6T 1Z1, Canada}
\affil[e]{Department of Physics, California Institute of Technology, Pasadena, California 91125, USA}
\affil[f]{Department of Physics, Stanford University, Stanford, California 94305, USA}
\affil[g]{Department of Physics, University of Cincinnati, Cincinnati, Ohio 45221, USA}
\affil[h]{Jet Propulsion Laboratory, Pasadena, California 91109, USA}
\affil[i]{Kavli Institute for Cosmological Physics, University of Chicago, Chicago, Illinois 60637, USA}
\affil[j]{Minnesota Institute for Astrophysics, University of Minnesota, Minneapolis, Minnesota, 55455, USA}
\affil[k]{National Institute of Standards and Technology, Boulder, Colorado 80305, USA}
\affil[l]{School of Physics and Astronomy, University of Minnesota, Minneapolis, Minnesota 55455, USA}
\affil[m]{Service des Basses Temp\'{e}ratures, Commissariat \`{a} l'Energie Atomique, 38054 Grenoble, France}
\affil[n]{Department of Physics, University of Illinois at Urbana-Champaign, Urbana, Illinois 61801, USA}
\affil[o]{Department of Astronomy, University of Illinois at Urbana-Champaign, Urbana, Illinois 61801, USA}
\affil[p]{Department of Physics, Harvard University, Cambridge, Massachusetts 02138, USA}
\affil[q]{Kavli Institute for the Physics and Mathematics of the Universe (WPI), UTIAS, The University of Tokyo, Kashiwa, Chiba 277-8583, Japan}
\affil[r]{Aix-Marseille Universit\'{e}, CNRS/IN2P3, CPPM, 13288 Marseille, France}
\affil[s]{Department of Physics, Enrico Fermi Institute, University of Chicago, Chicago, Illinois 60637, USA}

\authorinfo{Corresponding Author: James Cornelison\\E-mail: james\_cornelison@g.harvard.edu}

\begin{document} 
\maketitle
\begin{abstract}
The BICEP3 Polarimeter is a small aperture, refracting telescope, dedicated to the observation of the Cosmic Microwave Background (CMB) at 95GHz. It is designed to target degree angular scale polarization patterns, in particular the very-much-sought-after primordial B-mode signal, which is a unique signature of cosmic inflation. The polarized signal from the sky is reconstructed by differencing co-localized, orthogonally polarized superconducting Transition Edge Sensor (TES) bolometers. In this work, we present absolute measurements of the polarization response of the detectors for more than $\sim 800$ functioning detector pairs of the BICEP3 experiment, out of a total of $\sim 1000$. We use a specifically designed Rotating Polarized Source (RPS) to measure the polarization response at multiple source and telescope boresight rotation angles, to fully map the response over 360 degrees. We present here polarization properties extracted from on-site calibration data taken in January 2022. A similar calibration campaign was performed in 2018, but we found that our constraint was dominated by systematics on the level of $\sim0.5^\circ$. After a number of improvements to the calibration set-up, we are now able to report a significantly lower level of systematic contamination. In the future, such precise measurements will be used to constrain physics beyond the standard cosmological model, namely cosmic birefringence.
\end{abstract}


\section{INTRODUCTION}
\label{sec:intro}
In the standard model of particle physics, no interactions are known to cause parity violation in the electromagnetic interaction.
However, parity violations have been predicted and observed for the weak interaction\cite{1956Lee,1957Wu}.
Because electromagnetic and weak interactions are unified in the electroweak regime at very high energies, it has been proposed that the electromagnetic interaction may also violate parity.
In particular, cosmic birefringence is defined as the Universe being filled with a parity-violating field, interacting with the electromagnetic field.
One candidate is the interaction with a pseudo-scalar field through the Chern-Simons term in the Lagrangian density \cite{1988turner}.
Such pseudo-scalar fields include axion-like particles, which could be a signature for dark matter or dark energy\cite{2016marsh}.
Another mechanism that would give rise to cosmic birefringence is the interaction of photons with primordial magnetic fields at recombination \cite{2011caldwell}.
Any of these proposed mechanisms would manifest as a rotation of the linear polarization direction of photons, integrated along the line of sight.
Because CMB photons have traveled the further possible distance in the Universe, this is where the cosmic birefringence signal, if any, would be the strongest.

The CMB signal is an almost uniform black body spectrum at 2.725K which exhibits very small spatial anisotropies in temperature and polarization.
Temperature anisotropies probe density fluctuations in the early Universe, and have been mapped down to cosmic-variance limited precision by the satellite Planck \cite{2020planck}.
Polarization anisotropies are canonically described in terms of parity-even scalar E-modes and parity-odd pseudo-scalar B-modes. E-mode anisotropies map density fluctuations, while B-modes anisotropies map tensor fluctuations such as primordial gravitational waves. 
While E-mode anisotropies have been mapped with a high precision\cite{2020planck}, the detection of primordial B-modes would be a unique and strong case for inflationary models\cite{1997sph1}.
Cosmic birefringence, by rotating the polarization angle of CMB photons, is predicted to cause E $\rightarrow$ B leakage, which would otherwise not occur in the standard cosmological model.
This manifests as a non-zero signal in the EB and TB power spectra.
However, EB and TB signal can also be generated by instrumental systematics where the instrument polarization angle has been mis-calibrated\cite{2013Selfcal}.
For CMB observations whose focus is not to constrain cosmic birefringence, this effect is removed by minimising the EB and TB signal as correlations between T and B and between E and B are zero by construction\cite{2009wu}.
The resulting angle is then used as the instrument polarization angle for the rest of the analysis, hence the name of ``self-calibration" for this method \cite{2013Selfcal}.
The challenge to detect cosmic birefringence is therefore to precisely and accurately characterise the instrument polarization angle so that EB and TB signal from instrument polarization and from cosmic birefringence can be disentangled.
Many CMB experiments have reported progress on this over the past 20 years, as described in Table \ref{tab:bcconstraints}.

In this paper, we first introduce the BICEP3 Polarimeter in \S \ref{sec:b3}, and the Rotating Polarized Source (RPS) in \S \ref{sec:rps}. We then describe in details the calibration observations taken in January 2022 in \S \ref{sec:obs}, followed by an overview of our analysis pipeline and preliminary results in \S \ref{sec:analysis}. We conclude by outlining future steps to derive cosmic birefringence constraints from this data set in \S \ref{sec:conc}.

\begin{table}[ht]

    \centering
\begin{adjustbox}{width={\textwidth},totalheight={\textheight},keepaspectratio}
\begin{tabular}{ ccccc }

\hline
\hline
\textbf{Experiment} & \textbf{Frequency (GHz)} & \textbf{$\ell$ range} & \textbf{$\alpha (^\circ)\pm\text{(stat)}\pm\text{(sys)}$} & \textbf{Calibration Method}\\
\hline

\hline
\multirow{2}{*}{QUaD\cite{2009wu}} & 100 & \multirow{2}{*}{200-2000} & $-1.89\pm2.24 (\pm0.5)$ & \multirow{2}{*}{polarized source}\\
& 150 && $+0.83\pm0.94 (\pm0.5)$ &\\

\hline
BOOM03\cite{2009pagano} & 143 & 150-1000 & $-4.3\pm4.1$ & pre-flight polarized source \\

\hline
ACTPol\cite{2009finelli} & 146 & 500-2000 & $-0.2\pm0.5(-1.2)$ & ''As-Designed'' \\

\hline
WMAP7\cite{2011Komatsu} & 41+61+94 & 2-800 & $-1.1\pm1.4 (\pm1.5)$ & pre-launch polarized source / Tau A \\

\hline
BICEP2\cite{2013aiken} & 150 & 30-300 & $-1\pm0.2(\pm1.5) $ & Dielectric Sheet \\

\hline
\multirow{3}{*}{BICEP1\cite{2014selfcal}} & \multirow{3}{*}{100+150} & \multirow{3}{*}{30-300} & $-2.77\pm0.86(\pm1.3)$ & Dielectric Sheet\\ 
&&& $-1.71\pm0.86(\pm1.3)$ & Polarized Source \\
&&& $-1.08\pm0.86(\pm1.3)$ & ''As-designed'' \\

\hline
POLARBEAR\cite{2014polarbear} & 150 & 500-2100 & $-1.08\pm0.2 (\pm0.5)$ & Tau A\\

\hline
\textit{Planck}\cite{2016planckparity} & 30-353 & 100-1500 & $0.35\pm0.05 (\pm0.28)$ & pre-flight source / Tau A$^\dagger$\\
\hline
ACTPol\cite{2020choi} & 150 & 600-1800 & $-0.07\pm0.09$ & metrology+modeling+planet obs. \\

\hline
Planck (Minami et al.\cite{2020minami})& 30-153 & 100-1500 & $0.35\pm0.14$ & Planck data \\

\hline
Planck (Diego-Palazuelos et al\cite{2022Diego}) & 100-353 & 50-1500 & $0.30\pm0.11$ & Planck data \\

\hline
\hline
\end{tabular}
\end{adjustbox}
\caption{Uniform cosmic birefringence constraints from CMB experiments ordered chronologically by publication. Note that the two most recent results use a novel method not relying on the characterization of the instrument polarization response, but rather on galactic foregrounds properties\cite{2020minami,2022Diego}. However, this method suffers from other sources of systematic errors, in particular the incomplete knowledge of foreground polarization.}
{\footnotesize $^\dagger$ Calibration of polarization orientations was completed pre-flight in the near field and confirmed with celestial source Tau A within uncertainties \cite{2010planckgroundcal,2016planckspacecal}}.
\label{tab:bcconstraints}
\end{table}

\section{THE BICEP3 Polarimeter}
\label{sec:b3}
The BICEP3 CMB Polarimeter\cite{BKXV} is a 52cm, refracting telescope located at the geographic South Pole.
It is specifically designed to target the faint signal from primordial B-modes at degree angular scales.
It started scientific observation in 2016 with 2400 Transition Edge Sensor (TES) bolometers observing the sky at 95GHz.
The focal plane is divided into modular tiles that each contain $8 \times 8$ co-located and orthogonally polarized detector pairs.
To achieve a highly polarized response, the TES of each detector is coupled to a dual-slot antenna array that defines its individual polarization angle and polarization efficiency.
The telescope itself is mounted on a three-axis mount, that allows movement along the azimuth (Az) and elevation (El) axis, as well as around the boresight of the telescope, called deck rotation (Dk).
It is the rotation around this boresight axis that makes it possible to fully reconstruct the polarization signal from the sky, by allowing the polarized detectors to be sensitive to Q and U polarization components.

\begin{figure}
    \centering

    \begin{adjustbox}{width={\textwidth},totalheight={\textheight},keepaspectratio}
    \includegraphics{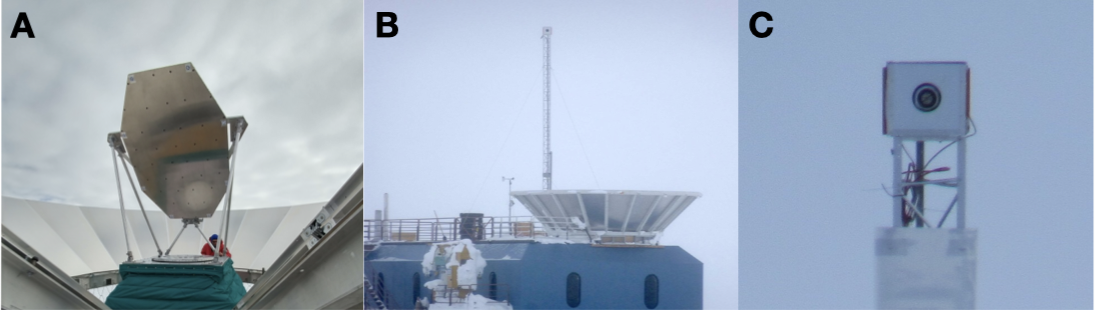}
    \end{adjustbox}
    \caption{Example images of the RPS calibration campaign.\textbf{A:} BICEP3 with FFF calibration mirror installed.\textbf{B:} RPS deployed atop a calibration mast, secured by guy-lines. \textbf{C:} Zoomed image of RPS installed in environmental enclosure shows red alignment strips which allow for a ($<1^\circ$) alignment with the telescope. Images courtesy of the BICEP Collaboration.}
    \label{fig:rpsimages}
\end{figure}

\begin{figure}
    \centering

    \begin{adjustbox}{width={\textwidth},totalheight={0.6\textheight},keepaspectratio}
    \includegraphics{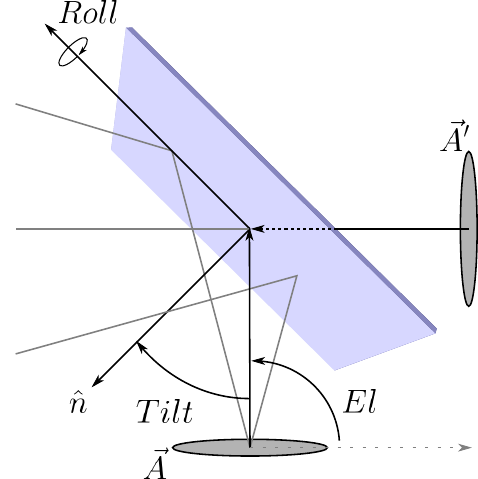}
    \end{adjustbox}
    \caption{Orientation of the FFF mirror with respect to the aperture. For a telescope pointing at zenith and a given azimuth, $Az$, and for \{$t=45^{\circ}$, $r=0^{\circ}$\}, the beam exiting the aperture $\vec{A}$ is redirected toward the horizon at $Az+180^\circ$ as $\vec{A'}$.}
    \label{fig:mirror_schematic}
\end{figure}

To perform field calibration activities with BICEP3, in particular beam calibration, we install calibration sources in the far field of the telescope by placing them atop a mast on a nearby building.
As-is, the receiver is however not able to look at sources below $50^\circ$ degrees in elevation.
The beams are thus redirected toward calibration sources (only a few degrees above the horizon) by using a monolithic aluminum honeycomb mirror.
The mirror is mounted directly atop the mount, such that it is comoving with the receiver in Az and El, but not in DK.
It is held in place by six aluminum struts and sits 1.4 m above the receiver window, as shown in panel A of Figure \ref{fig:rpsimages}.
We define the orientation of the mirror using two axes that form an orthonormal basis with the vector normal to the plane of the mirror (Figure \ref{fig:mirror_schematic}). 
The tilt axis ($t$) is defined in parallel to the elevation axis but of opposite handedness such that a positive tilt reflects the beams ``behind" the telescope.
The roll axis ($r$) is the cross between the mirror normal and the tilt axis.
The target parameters for the mirror position are $t=45^{\circ}$ and $r=0^{\circ}$.
The determination of the precise position of the mirror is of importance to properly characterise the pointing model of the instrument when the mirror is installed, and their determination is detailed in \S \ref{sec:analysis}.
The RPS itself is installed atop a 12-meter mast on the Martin A. Pomerantz Observatory (MAPO), located $\sim200$m away from the receiver, as shown in panel B and C of Figure \ref{fig:rpsimages}.

\section{The Rotating Polarized Source}\label{sec:rps}
\begin{figure}[ht!]
    \centering

    \begin{adjustbox}{width={\textwidth},totalheight={\textheight},keepaspectratio}
    \includegraphics{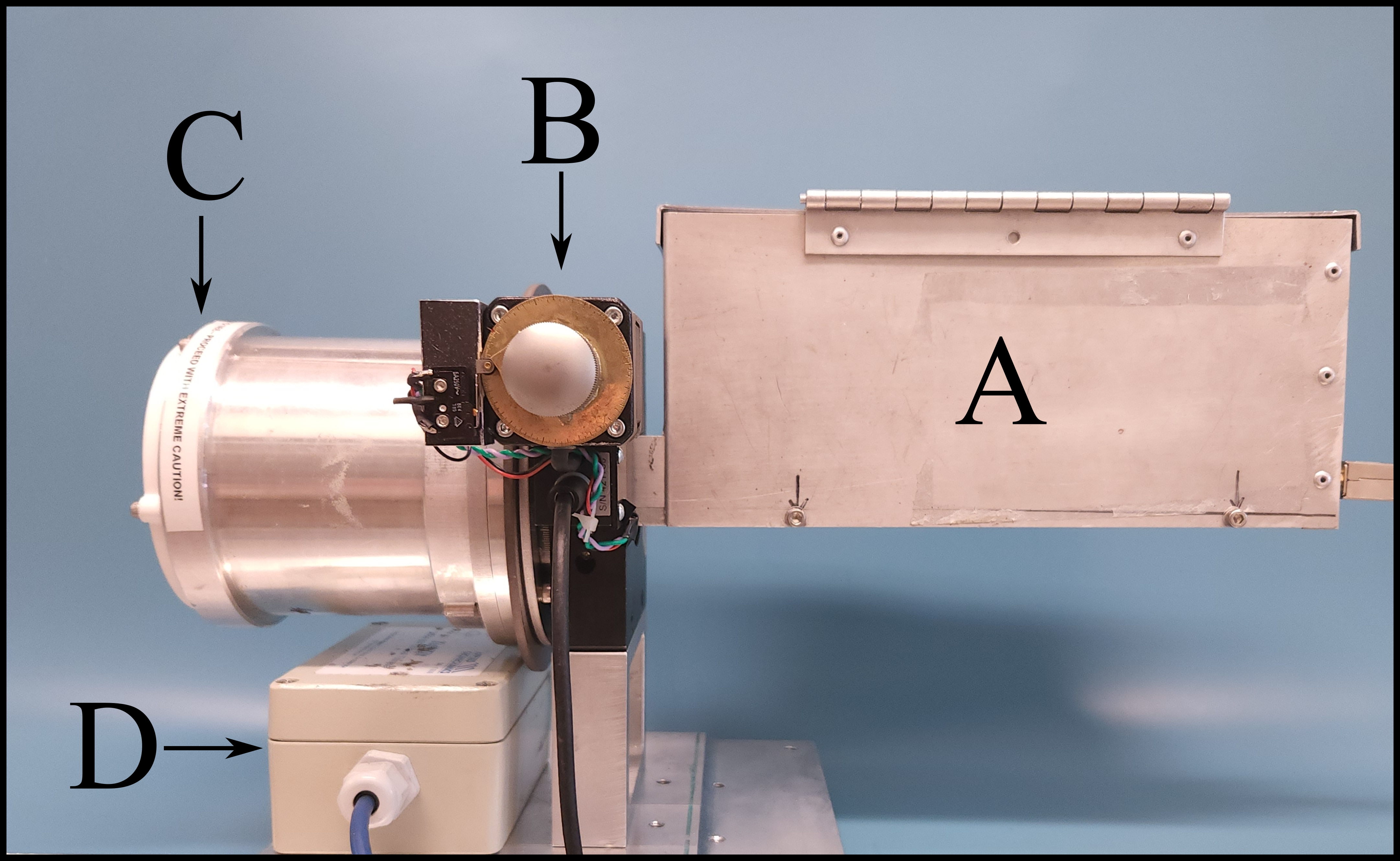}
    \end{adjustbox}
    \caption{Diagram of the Rotating Polarized Source. An electrically chopped, quasi-thermal noise source (\textbf{A}) with an instantaneous, 10GHz spectrum centered at 95GHz is fixed to an optical rotation stage (\textbf{B}) such that the rectangular feedhorn of the source is colinear with the stage's axis of rotation. We place a wire grid (\textbf{C}) in front of the feedhorn to further polarize the output. We calibrate and monitor the orientation of the wire grid with respect to gravity using a high-precision tilt meter (\textbf{D}).}
    \label{fig:rpshardware}
\end{figure}

We use a human-made, ground-based calibration source for the purpose of characterizing the polarization properties of the BICEP3 detectors, called the Rotating Polarized Source (RPS).
In summary, the RPS comprises a Broad Spectrum Noise Source (BSNS) with a linearly polarized output which is polarized further with a wire grid polarizer (WGP) -- both of which are fixed to an optical rotation stage as seen in Figure \ref{fig:rpshardware}.
The orientation of the wire grid is calibrated with respect to gravity with a high-precision tilt meter which is fixed to the same aluminum plate as the rotation stage.

The BSNS is a quasi-thermal, electrically chopped noise source with a 10GHz instantaneous band centered at 95 GHz.
The RF chain of the BSNS starts by amplifying the Johnson noise of a $50\Omega$ termination by 70dB using two 12-to-20 GHz broadband power amplifiers.
To aid with background rejection, we insert a pin switch after the power amplifiers and chop the RPS output electrically with a 20 Hz square wave with a 50\% duty cycle.
The reference square wave is fed into the telescope data acquisition system\cite[\S 7]{BKXV} in the neighboring building via low-loss fiber optic cabling for use during analysis.
The low-frequency signal is then put through a frequency multiplier to bring it to our desired frequency range (in this case a sextupler).
A single WR-10 power amplifier is used to amplify by another 10dB.
We also installed two WR-10 waveguide attenuators to provide over 60dB of dynamic range on the output amplitude.
Finally, we run the signal through a 90-to-100 GHz bandpass filter before the signal exits from a rectangular feedhorn.
The resulting signal coming from the source is instantaneously broadband over 10GHz with a flatness of $\sim1$ dB and has a measured output power of 2.3 dBm or an effective noise temperature of $\sim 12$ billion Kelvin.
This design was chosen over other options such as Gunn Diode Oscillators in order to sample the entirety of our beam in frequency space.

The polarizing wire grid is sourced from Microtech \cite{2022microtech} and is made of $15\,\mu m$ tungsten wires with $80\,\mu m$ spacing.
The variance on the spacing differs by less than the width of a wire.
The cross-polar power of the wire grid has been measured to be below 0.2\%.

The rotation stage is a Thorlabs NR360S precision optical stage with a homing precision of $0.06^\circ$ and a bi-directional repeatability of $0.002^\circ$ \cite{2022thorlabs}.
We used the rotation stage as specified in previous years and found the quoted homing precision of $0.06^\circ$ to be sufficient.
This year however, we installed an external homing switch on the knob of the stepper motor in parallel to the internal homing switch which improved the homing repeatability to $<0.01^\circ$.
When the BSNS is attached to the rotation stage, we ensure the waveguide of the source is collimated with the rotation axis of the stage using a room-temperature detector diode and recording the amplitude modulation as the source rotates in and out of polarization alignment such that the peaks of the modulation curve differ by $<1\%$.

We use an Applied Geomechanics Tuff Tilt precision tilt meter\footnote{This company has since gone out of business} as our primary means of connecting the orientation of the WGP to gravity.
With the stage at its home position, we measure the orientation of the wires with respect to a chosen reference plane on the rotation stage itself to $<0.01^\circ$.
After assembling the RPS as it looks in Figure \ref{fig:rpshardware}, we level that reference plane using a precision machinist level and record the output of the tilt meter to a precision of $\sim0.01^\circ$.
Thus, when deploying the RPS for observations, we can repeatably align the WGP with respect to gravity by ensuring the tilt meter reads that same initial value.
We note that the tilt calibration does change when moving from room temperature to the $-30^\circ C$ Antarctic climate, but have confirmed that this change is consistent within manufacturer specifications ($\sim 1\text{arcsec}/{}^\circ C$).

In our 2018 calibration campaign, we encountered an issue in the tilt meter electronics which dominated the systematics of that measurement.
During that year, the shielded cabling that was used for all BICEP calibration equipment was replaced with cables with more conductors but less shielding.
As a result, we experienced abnormally high capacitive coupling between the stepper motor of the rotation stage and the tilt meter readout which in turn caused an apparent shift in the tilt meter reading whenever the stepper motor stopped at a new position.
Because we used the same tilt meter calibration described above, the reference of the WGP to gravity was lost and our best constraint on the absolute angle due to this issue was raised to $\sim0.5^\circ$.
During our calibration campaign this year, we mitigated the electrical coupling between the tilt meter and rotation stage by connecting them separately with their own twisted-shield-pair cables and changing the readout mode of the tilt meter from single-ended to differential.
This improved the uncertainty on the tilt meter calibration from $0.5^\circ$ to $0.01^\circ$.

\section{Observations}
\label{sec:obs}

\subsection{Moon Observations}
\label{moon_obs}
We observe the Moon in order to derive the position and orientation of the monolithic calibration mirror mentioned in \S \ref{sec:b3} and to control systematics originating from pointing errors.
We start from the principle that we know very precisely the pointing model of our mount from optical starpointing without the mirror on one hand, and the ephemeris of the moon from the JPL Horizons database \cite{2022Horizons} on the other hand. We can therefore fit for the mirror tilt and roll in the pointing model by minimizing the difference between beam centers derived from the CMB (without the mirror) and the beam centers derived from Moon observations with the mirror installed.
We amounted $\sim70$ hours of moon data divided roughly into 24 hour periods with one period on 3 December 2021, the second on 9 December 2021, and the final period occurring after completion of the RPS observations on 25 January 2022, to track the stability of mirror parameters over time.

\subsection{RPS Observations}
\begin{figure}
    \centering

    \begin{adjustbox}{width={\textwidth},totalheight={\textheight},keepaspectratio}
    \includegraphics{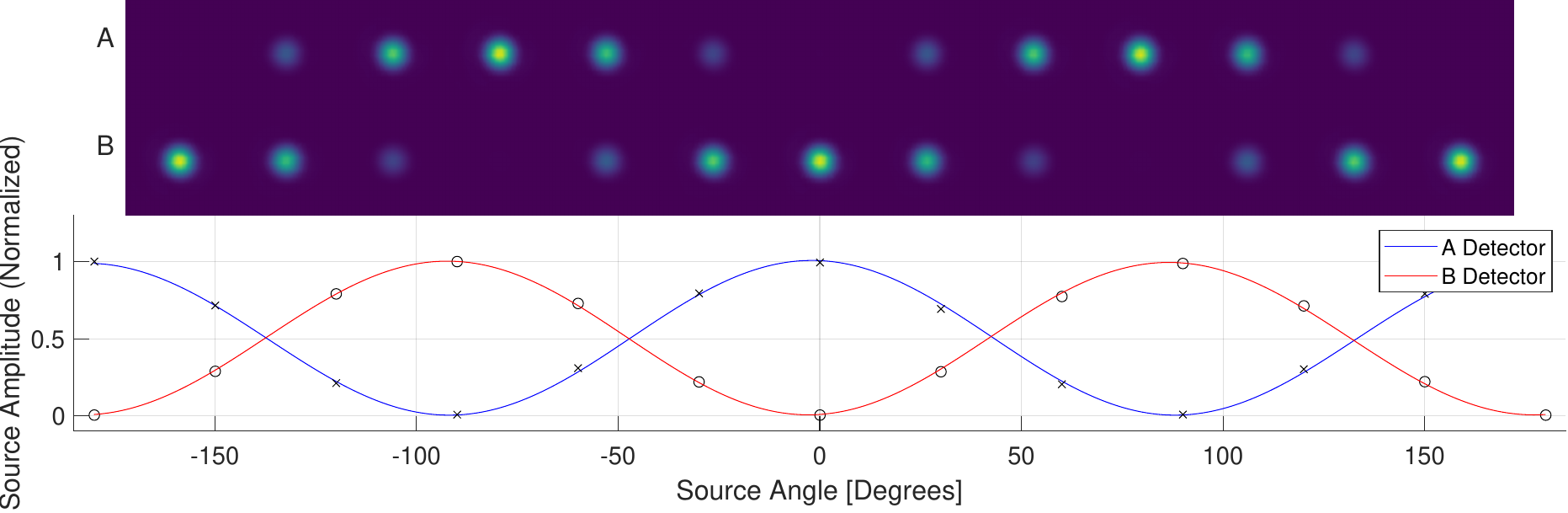}
    \end{adjustbox}
    \caption{(\textbf{Top}).Binned and smoothed RPS timestreams for A and B
detectors showing beams.  (\textbf{Bottom}) Resulting modulation curve data
(\textbf{x/o’s}) and fits to the model (\textbf{lines}).}
    \label{fig:modcurve}
\end{figure}

Observations of the RPS were conducted between 24 December 2021 and 28 January 2022.
A typical RPS observation consists of thirteen $9^\circ\text{Az}\times2^\circ\text{El}$ rasters intended to map beams for $\sim20$ detector pairs on the focal plane at a time.
The RPS is commanded to a selected polarization angle and is fixed at that position during a given raster.
When that raster is complete, the RPS is commanded to a new position before the next raster begins.
The RPS is rotated over $360^\circ$ from $-180^\circ$ to $180^\circ$ with respect to gravity in $30^\circ$ increments.
The amplitude of the beams modulate as the polarization orientation of the RPS moves in an out of alignment with a given detector, producing a modulation curve.
Figure \ref{fig:modcurve} shows beam maps and resulting modulation curves for a detector pair from a typical RPS observation.

We also conducted a number of cross-check observations where a single parameter in the observing schedules was adjusted while keeping all other parameters the same. 
The schedules used and their purposes are listed in Table \ref{tab:crosschecks}. Standard RPS observations and Moon observations are type 5 and 7, respectively. Type 8 and various type 9 are quick observations to keep track of system stability throughout the calibration campaign. Types 6, 10, and 11 have been designed to address specific systematic error concerns, and their analysis is detailed in \S \ref{sec:crosschecks}.
In total, we were able to collect over two straight weeks of standard RPS observations and one full week of systematic cross-check observations.

\begin{table}[htbp]

    \centering
\begin{adjustbox}{width={\textwidth},totalheight={\textheight},keepaspectratio}
\begin{tabular}{ ccccc }

\hline
\hline
\textbf{Schedule Type} & \textbf{Obs. Length (hrs.)} & \textbf{Total Time (hrs.)} & \textbf{Adjusted Parameter} & \textbf{Systematics Probed} \\
\hline

\hline

\hline
\cellcolor{purple!20}
5 & 13.5 & 420.4 & None & Standard RPS Observations \\
\cellcolor{teal!20}
6 & 13.6 & 13.6 & RPS angle -180 to 180 minus DK angle & Importance of nulling mod curve \\
\cellcolor{purple!20}
7 & 1 & 70.7 & Target Moon, only one raster & Standard Moon Observations \\
\cellcolor{orange!20}
8 & 0.1 & 0.5 & No mount movement & Rotation stage operation checks \\
\cellcolor{orange!20}
9 & 0.5 & 22.9 & quick-turnaround $1^\circ\times1^\circ$ single-pair map  & System stability \\
\cellcolor{orange!20}
9.1 & 0.3 & 3.5 & $1^\circ\times1^\circ$ map, $45^\circ$ increments & $\sigma$-Angle vs. Samples \\
\cellcolor{orange!20}
9.2 & 1.1 & 11.0 & $1^\circ\times1^\circ$ map, $10^\circ$ increments & $\sigma$-Angle vs. Samples \\
\cellcolor{orange!20}
9.3 & 0.5 & 4.6 & $1^\circ\times1^\circ$ map, No Homing between obs. & RPS homing precision \\
\cellcolor{teal!20}
10 & 3.1 & 9.2 & RPS fixed \& boresight rotated instead & Pointing model validation \\
\cellcolor{teal!20}
11 & 8.5 & 46.4 & Rasters $27^\circ$ in Az & Short- \& long-timescale mirror deformation \\
\hline
\hline
\end{tabular}
\end{adjustbox}
\caption{Suite of observation schedules used in this dataset. Regular observations are shown in {\color{purple}red}, stability cross-checks in {\color{orange}orange} and systematics cross-checks in {\color{teal}green}.}
\label{tab:crosschecks}
\end{table}

\section{Analysis and Results}
\label{sec:analysis}

\subsection{Data Reduction}
\begin{figure}
    \centering
    \begin{adjustbox}{width={\textwidth},totalheight={\textheight},keepaspectratio}
    \includegraphics{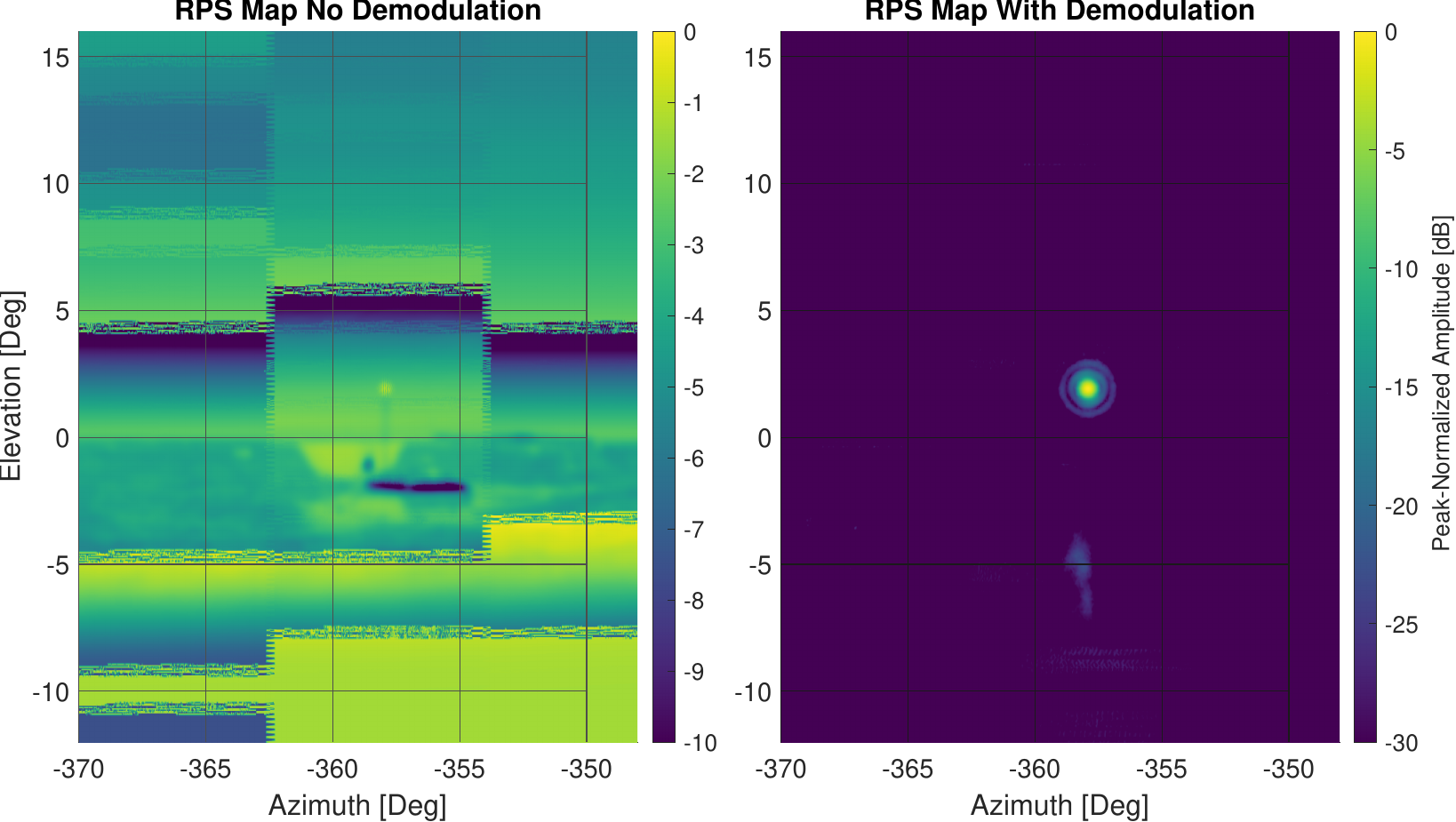}
    \end{adjustbox}
    \caption{A comparison of a beam map from a single detector with and without square-wave demodulation. (\textbf{Left}) The raw signal includes both the source signal ($-358^\circ$ Az / $2^\circ$ El) and all ambient signals around it (the BICEP Array groundshield can be made out just below the source), many of which are at similar apparent temperatures. (\textbf{Right}) The demodulated signal rejects all ambient signals to $<30$ dB such that the main beam and Airy rings are easily seen. The feature in the demodulated map below the source at $-358^\circ$ Az / $-5^\circ$ El is a real reflection of errant source power off the snow or tilted BICEP3 groundshield.}
    \label{fig:demod}
\end{figure}

We perform low-level data reduction in the form of transfer function correction \cite[\S 13]{BKII} and square-wave demodulation.
The square-wave demodulation is performed via software in a manner similar to a physical lock-in amplifier. We mix the detector timestreams with the reference signal, low-pass filter to isolate the DC power, and then downsample to the new Nyquist frequency.
The reference signal is a 20 Hz square wave with a 50\% duty cycle and we apply a 19.5 Hz to 20.5 Hz bandpass filter to the reference signal to eliminate the higher-order harmonics and thus improve background rejection.
Figure \ref{fig:demod} shows the effectiveness of the demodulation function -- typical SNR for a $625\,\text{deg-sq.}$ map is $\mathcal{O}(6000)$.

We perform quality cuts to the data throughout the analysis using the same cutting approach as in \cite[\S 10]{BKXV} where applicable. We also make manual cuts on the per-detector fit parameters described below where poor fits are the obvious cause, e.g. detector beam centroids that are several degrees away from the median.

\subsection{Parameter Estimation}\label{params}

\begin{figure}
    \centering
    \begin{adjustbox}{width={\textwidth},totalheight={0.4\textheight},keepaspectratio}
    \includegraphics{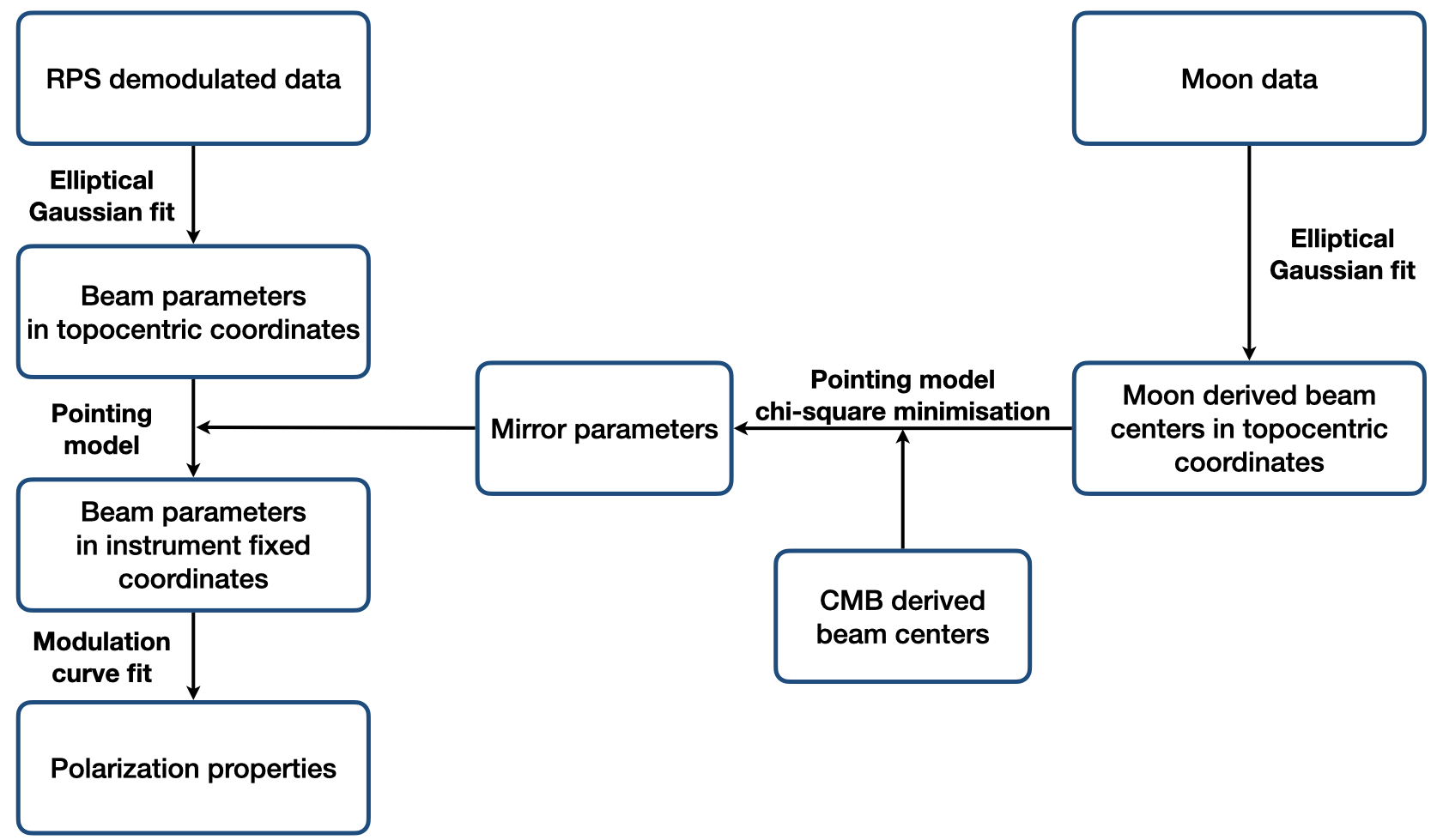}
    \end{adjustbox}
    \caption{Simplified analysis flowchart}
    \label{fig:flowchart}
\end{figure}

In this section, we perform several successive fits to extract relevant parameters. We start by deriving mirror parameters using the Moon as reference (as outlined in \S \ref{moon_obs}). We then use these parameters as part of the analysis of the RPS modulation curve to derive polarization angle properties. A simplified flowchart of the analysis pipeline is shown in Figure \ref{fig:flowchart}.

\subsubsection{Mirror Orientation}
We first perform a fit of elliptical Gaussian profiles to beam maps of the Moon in topocentric horizontal coordinates:
\begin{equation}
    B(\mathbf{x}) = A e^{(\mathbf{x}-\mathbf{\mu})^T\Sigma^{-1}(\mathbf{x}-\mathbf{\mu})}
\end{equation}
Where $A$ is the amplitude, $\mathbf{\mu} = (Az_0\,\,El_0)$ is the beam centroid, and

\begin{equation}
\Sigma =
\begin{bmatrix} \sigma_{Az}^2& \rho \sigma_{Az} \sigma_{El} \\ \rho \sigma_{Az} \sigma_{El} & \sigma_{El}^2 \end{bmatrix}
\end{equation}

where $(\sigma_{Az},\,\sigma_{El})$ is the beam width, and $\rho$ is the beam ellipticity. This first step does not require knowledge of the mirror position, and we extract from it the positions of the beam centers ($Az_0,\,El_0$).

We then use a pointing model to convert the Moon-derived beam centroids from topocentric horizontal coordinates to an instrument fixed coordinate system $(x,\,y)$\footnote{For more information on our instrument-fixed coordinate system, see \cite[\S 3]{BKXI}.}.
The pointing model ($PM$) is parameterized by the mirror orientation, tilt ($t$) and roll ($r$), and we use instrument-fixed beam centroids derived from the CMB without a mirror, $\mathbf{x}_{CMB} = (x_{CMB},y_{CMB})$, as our fiducial data in our chi-square:

\begin{equation}
    \chi^2 = \sum\frac{\left(\mathbf{x}_{CMB}-PM\left(t,r|Az_0,\,El_0,\,Mount,\,Moon\right)\right)^2}{\sigma^2}
    \label{eq:chi2}
\end{equation}

where $Mount$ and $Moon$ are inputs into the pointing model which describe the location and orientation of the mount axes, and the on-sky orientation of the Moon respectively.
We minimize Eq. \ref{eq:chi2} with respect to $t$ and $r$ and use the result of this fit as inputs to the pointing model in the following analysis.

\subsubsection{Polarization Properties}
We fit elliptical Gaussian profiles using the same method above except for RPS observations we fit across all beams in a rasterset simultaneously where the beam amplitude is a free parameter for each raster but the beam center, beam width, and beam ellipticity are single parameters fit across the entire rasterset.
The choice to fit a single Gaussian profile across all beams is to mitigate erroneous fits when the RPS and detector are $90^\circ$ out of phase and the constraining power on the beam is poor. The resulting amplitudes produce a modulation curve as a function of source angle $\theta$.

We model the polarization response $A$ for a given detector as:
\begin{equation}\label{eq:modcurve}
    A(\theta) = G\left(\cos\left(2\left(\theta+\psi\right)\right)-\frac{\epsilon+1}{\epsilon-1}\right)\left(n_1\cos(\theta)+n_2\sin(\theta)+1\right)+N
\end{equation}
where $G$ is the detector gain, $\psi$ is the detector polarization angle with respect to the source, $\epsilon$ is the detector cross-polar response, ($n_1,\,n_2$) are nuisance parameters to account for miscollimation of the source rotation axis, and N is our noise model. We fit for polarisation angles $\psi$ and polarisation efficiencies $\epsilon$.

The polarization angle $\psi$ in Equation \ref{eq:modcurve} is a relative measurement and is agnostic to the orientation of the focal plane. 
We define an angle $\phi_d$, which is the polarization angle of a detector with respect to the focal plane, by combining $\psi$ with the angle between the telescope's focal plane orientation and the RPS's orientation, called $\phi_s$, which is computed through a pointing model.

\begin{equation}
    \phi_d \equiv \psi + \phi_s
\end{equation}

Finally, we compute the pair-differenced polarization parameters of a given detector pair via the amplitude of their response, $S$, to Stokes parameters
Q and U\footnote{Assuming the gain of A and B detectors are equal.}:

\begin{equation}
    \begin{split}
        &S_{Q} = \frac{S_{+Q}-S_{-Q}}{2} =  \frac{\cos(2\phi_A)-\cos(2\phi_B)}{2}\\
        &\\
        &S_{U} = \frac{S_{+U}-S_{-U}}{2} =\frac{\sin(2\phi_A)-\sin(2\phi_B)}{2}\\
    \end{split},
\end{equation}

where $\phi$ is implicitly $\phi_d$.

The pair-difference polarization angle $\phi_{pair}$ is 
$\epsilon_{pair}$ is:
\begin{equation}\label{eq:phipair}
\begin{split}
    &\phi_{pair} = \frac{1}{2}\tan^{-1}\frac{S_U}{S_Q}\\
\end{split}
\end{equation}

\subsection{Polarization Angles}
\begin{figure}
    \begin{subfigure}{.49\textwidth}
    \centering
    \includegraphics[width=1\linewidth]{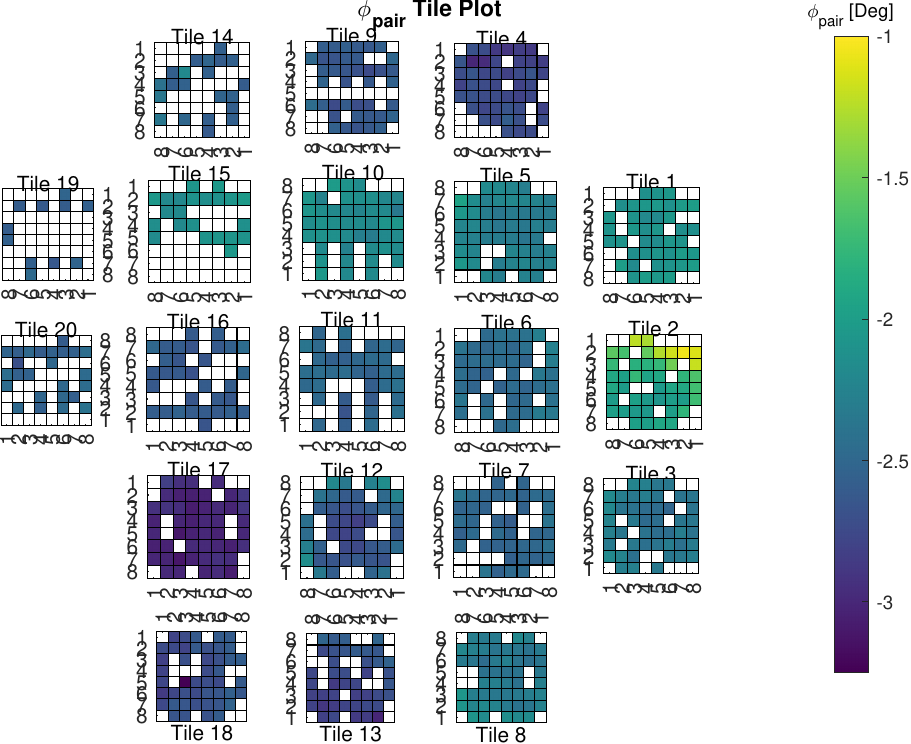}
    \end{subfigure}
    \begin{subfigure}{.49\textwidth}
    \centering
    \includegraphics[width=1\linewidth]{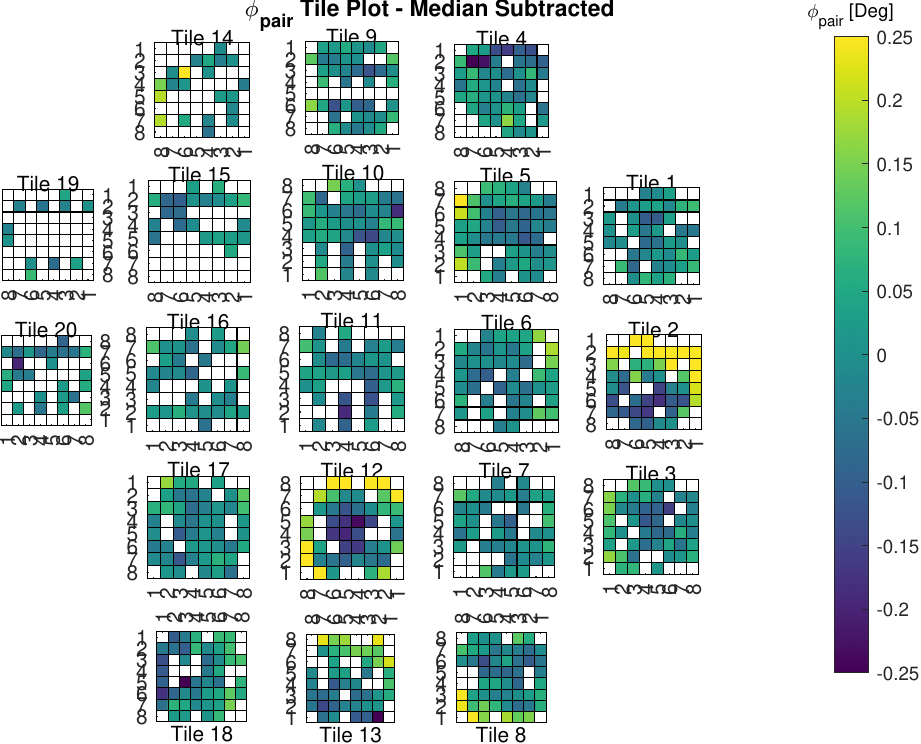}
    \end{subfigure}
\caption{Tile plots of per-pair polarization angles for the 2022 RPS calibration season. Each square represents a detector pair's location relative to its physical location on the focal plane. Each $8\times8$ group of squares represents a modular tile with the surrounding numbers denoting the column and row for readout purposes. (\textbf{Left}) The color scale represents the per-pair polarization angle referenced to the physical orientation of the focal plane. We see a $\sim-2.5^\circ$ global rotation with tile-to-tile shifts in polarization angle dominating the structure across the focal plane. (\textbf{Right}) The color scale of this plot shows the polarization angles with per-tile-medians subtracted which allows tile-specific structure to be seen.
}\label{fig:polangles}
\end{figure}

We present the RPS-derived polarization angles in Figure \ref{fig:polangles}.
Consistent with our 2018 results \cite{2020cornelison}, we found that individual tiles differ slightly in overall rotation with respect to one another with an overall scatter of $0.28^\circ$ and a smaller scatter within the tiles themselves with the median standard deviation of $0.065^\circ$.
Some tiles possess systematic features (tiles 2 and 12 in Figure \ref{fig:polangles}, for instance) but a majority are roughly Gaussian in distribution.
We calculate a global polarization angle of $-2.45^\circ$ with respect to gravity, calculated as a uniform mean across all viable pairs and across all observations.
It should be noted that per-pair polarization angles interface with our CMB data in a way that considers their individual sensitivities and coverage of the sky, so the actual effective global polarization angle is expected to differ slightly and our reported value should be considered preliminary.

\subsection{Comparing 2018 and 2022 data}\label{sub:results}
\begin{figure}
    \centering
    \begin{adjustbox}{width={0.9\textwidth},totalheight={\textheight},keepaspectratio}
    \includegraphics{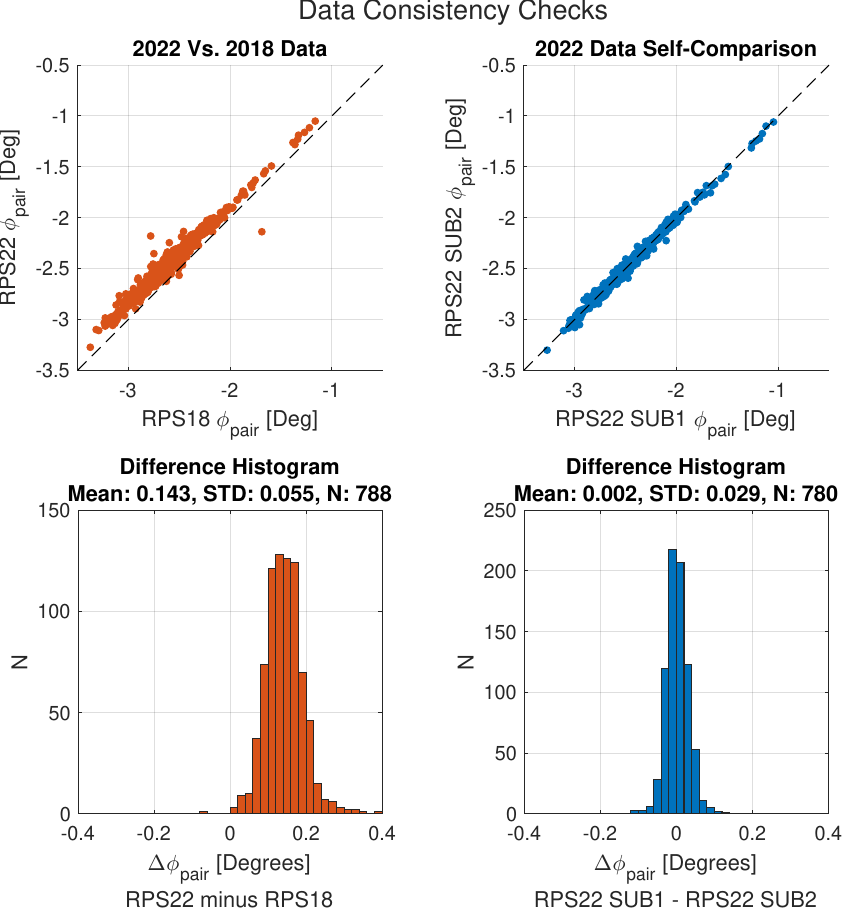}
    \end{adjustbox}
    \caption{A comparison of 2018 and 2022 RPS results as well as a self-consistency check of the 2022 RPS data. Each point in the \textbf{scatter plots} represents the polarization angle of a detector pair which has been averaged across a given set of data. In the \textbf{histograms}, the two averaged polarization angles of a given detector pair are subtracted and binned. {\color{matred}\textbf{Left} column}: The x-axis is the averages of the complete 2018 dataset (five type-5 observations at four unique DK angles) and the y-axis is the averages over all available 2022 data (ten type-5's, one type-6, and five type-11's). {\color{matblue}\textbf{Right} column}: the x- and y-axes are subsets of available 2022 data which were evenly split across all types and DK angles.}
    \label{fig:2022_2018_compare}
\end{figure}

\begin{figure}
    \centering
    \begin{adjustbox}{width={0.9\textwidth},totalheight={\textheight},keepaspectratio}
    \includegraphics{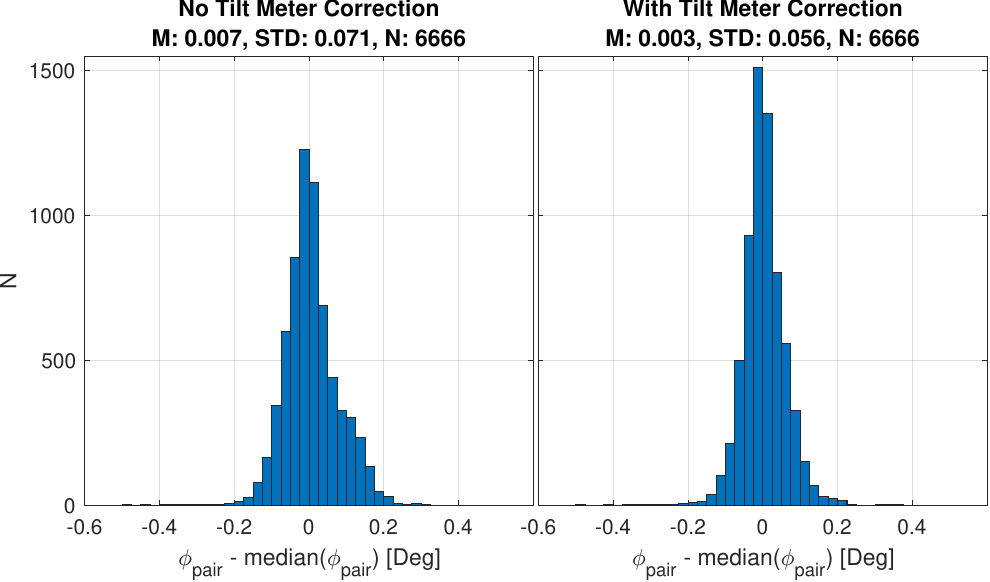}
    \end{adjustbox}
    \caption{A comparison of estimated polarization angles before (\textbf{Left}) and after (\textbf{Right}) tilt correction. We subtract from detector pair the median for that pair over all standard science observations (Type 5 in Table \ref{tab:crosschecks}) which allows for direct comparison of pairs while conserving the intrinsic scatter. We find that correcting for tilt improves reduces the scatter by $0.015^\circ$.}
    \label{fig:tilt_compare}
\end{figure}

We compare the per-pair polarization angles between the dataset taken in 2018 to our recent 2022 dataset in Figure \ref{fig:2022_2018_compare}. 
We find an offset in the overall polarization angle of $0.14^\circ$ which is due to the failure in the tilt meter calibration in 2018.
Additionally, when we split the data into two subsets distributed roughly evenly between the boresight angles at which they were taken, we find the data to be self-consistent with a scatter of $0.03^\circ$ representing the combination of statistical and systematic uncertainty.
The increased scatter when comparing the 2018 to 2022 data, we suspect, is in part due to the lack of tilt correction in the 2018 data as evidenced by Figure \ref{fig:tilt_compare}.

\subsection{Additional Cross Checks}
\label{sec:crosschecks}
In this section, we detail the motivation behind systematics cross-checks, and preliminary results from these observations.

\subsubsection{Mirror systematics}
\begin{figure}
    \centering

    \begin{adjustbox}{width={0.9\textwidth},totalheight={\textheight},keepaspectratio}
    \includegraphics{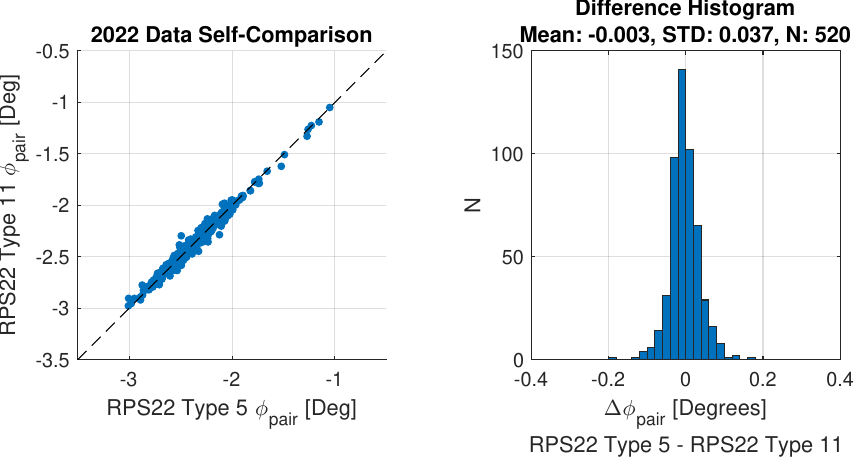}
    \end{adjustbox}
    \caption{A comparison between standard observations and observations which sampled more of the focal plane per rasterset. Each point in the \textbf{left scatter plot} represents the polarization angle of a detector pair which has been averaged across all type-5 and type-11 observations for the y-axis and x-axis respectively. In the \textbf{right histogram}, the two averaged polarization angles of a given detector pair are subtracted and binned. We conclude from these plots that any variations due to short- or long-timescale drifts average down to the same value.}
    \label{fig:t5vst11}
\end{figure}

A primary source of possible systematics comes from the FFF mirror. We identified four possible sources of mirror systematics:

\begin{itemize}
    \item Deformation of the mirror due to gravity (long time scale) or to diurnal fluctuation due to the position of the Sun and/or the weather (shorter time scale): to address this, we performed cross-checks in-between each RPS schedule (type 9, see Table \ref{tab:crosschecks}) looking at only a few detector pairs to make sure the polarization properties were not drifting over time. We used these as sanity checks when in the field, and a more thorough analysis of this data is on-going. We also designed a specific type of schedule (type 11) where we took rastersets at three azimuth offsets (-13, 4.5, 4 degrees), for a limited elevation range. This allowed us to acquire beams for detectors on opposite sides of the focal plane for a given rasterset. Effectively, more of the focal plane is sampled on a shorter time scale, which mitigates long term trends. By comparing angles derived from of type 11 vs type 5 (standard) schedules in Figure \ref{fig:t5vst11}, we show that variations on both of the timescales average down to the same value.
    \item Imperfect flatness on the mirror: this can affect the apparent location of beam centers redirected by the mirror. Using data taken at different boresight angles, we have shown that the error between instrument-fixed beam centroids derived from the CMB used as reference and RPS-derived beam centers (see \S\ref{params}) is uncorrelated with where a given beam reflects on the mirror. This indicates that imperfect flatness of the mirror is a subdominant cause of systematic error.
    \item Differential polarization of aluminum: we account for refractive properties of the aluminum mirror by calculating the reflection loss via Fresnel's equations. Preliminary results show that A and B detectors can get reflected by a maximum of $\sim 0.1^{\circ}$ in equal and opposite directions depending on the angle of incidence on the mirror which shows agreement with our initial models. Fortunately, any effects on the polarization angle from differential reflection cancels when computing the pair-difference signal, so no significant impact on the systematics is expected.
    \item Uncertainties in the pointing model: as we already mentioned, the characterization of the parameters of the pointing model, in particular the mirror position, is crucial to ensure an accurate reconstruction of the detectors beam centers and polarization properties. In order to mitigate systematics coming from the relative orientation of the RPS with respect to the mirror, we took data by keeping the RPS fixed and rotating the telescope around its boresight (type 10 schedules). However, the scanning strategy that we implemented for this test did not properly account for this boresight rotation in the choice of az/el ranges. We therefore do not possess enough data for the majority of the detectors as the focal plane coverage is very partial, and cannot form any conclusions from this test. 
\end{itemize}

To summarize, systematics mostly affects the determination of the beam centers as described in \S \ref{params}. This will result in residuals between the CMB-derived beam centroids, that we use as reference, and the RPS-derived beam centers that we are estimating which in turn affects how the polarization from the RPS is projected onto the telescope. Our approach is to model these residuals as resulting from a global rotation and scaling of the CMB-derived position. By doing this, we acknowledge that they are some unknowns in the mirror position, and we try to account for them in a parameter-blind way. This allows us to reduce the uncertainty on final polarization angle as projected onto the sky when observing the CMB. We currently find, on average, a global rotation of the focal plane to be $<0.01^\circ$ and a scaling of $<0.5\%$.

\subsubsection{Modulation curve}
The modulation curve produced by the coupling of the RPS with a given detector is simply defined by a cosine-squared function under ideal conditions. 
However, we consider a couple of factors that can introduce errors in our analysis:

\begin{itemize}
\item Polarization projection: pointing misaligment or rotation miscollimation can affect how the orientation of the source projects onto the focal plane which in turn affects the shape of modulation curve in a complicated way.
Further, while a complicated pointing model can feasibly be constructed which accounts for such non-idealities, it is not always guaranteed that the modulation curve is sufficiently sampled to constrain extra nuisance parameters.
As indicated in \S \ref{sec:rps}, we take great measures during our calibration campaigns to ensure that misalignment and collimation effects are mitigated as much as possible.
However, we are approaching the level of precision where even small pointing errors might introduce significant bias in our analysis.
As such, we acquired some short-duration type-9 observations of a single detector pair but with roughly $\sim40$ samples of the modulation curve (compared to our standard 13) to examine how capturing more information about the curve's shape might improve our control over systematics.
At this moment, these data are still being processed.
\item Curve nulling: while the shape of the modulation curve constrains pointing errors, sampling at the peaks and nulls of the curve can marginally improve polarization angle constraints and greatly improve constraint on the cross-polar response.
During previous years, the RPS would rotate over $360^\circ$ from $-180^\circ$ to $180^\circ$ with respect to gravity regardless of the DK at which we were observing.
In this year's campaign, we intended to include an additional offset to the source orientation such that the peaks and nulls of the modulation curve was captured at every DK that was observed.
However, this feature was excluded in error and was not discovered until late into the season. All data far away from DK's of $0^\circ$ and $90^\circ$ do not appropriately capture the nulls and peaks as intended.
Upon the discovery of this omission, we created and captured a single type-6 observation for this purpose which observed at a DK of $135^\circ$.
Currently, we find no change in the uncertainty between this observation and others at the same DK angle, implying that other sources of uncertainty are still dominant.
\end{itemize}

\section{CONCLUSIONS \& FUTURE STEPS}
\label{sec:conc}

We took data to measure BICEP3 detector's polarization properties by using a calibration mirror to observe a Rotated Polarized Source (RPS) atop a mast.
Careful characterization and improvements of the measurement setup have been performed since the last calibration run in 2018, allowing for a significant reduction in systematic contamination.
We took great care in precisely and accurately characterising the mirror position parameters that are key in reconstructing individual detector beam centroids and polarization properties.
In particular, building on the experience from previous campaigns, we designed specific cross-checks to improve our systematic control.
The analysis of this data is still underway, but preliminary results indicate that systematic effects are well controlled. 

We have reconstructed the per-pair polarization angles for more than 80\% of working detector pairs, reporting a global instrument polarization angle of $-2.45^\circ$.
This data set is also consistent with data taken in 2018 to the limit that the 2018 data was affected by previous systematics.
We also report a high level of self-consistency in the data across boresight angle rotation and schedule types, with a scatter of $0.03^\circ$ comprising the combination of systematic and statistical uncertainties thus far.
Our current work is focused primarily on the finalization of our systematics analysis which probes potential sources of error due to non-idealities in the calibration mirror or optical coupling between the telescope and calibration source.

Our immediate next steps will be to use this data to set constraints on signals of cosmic birefringence by integrating as-measured polarization properties into the BICEP/Keck CMB analysis pipeline \cite{BKXIII}.
The final constraints will be completed with a multi-component analysis that takes into account foreground complexity, in particular the effect of foregrounds on EB/TB power spectra.
Given our current efforts and preliminary results, we are optimistic that this analysis will provide the tightest constraints on cosmic birefringence to-date.

\section{ACKNOWLEDGEMENTS}
The BICEP/\textit{Keck} project (including BICEP2, BICEP3 and BICEP Array) have been made possible through a series of grants from the National Science Foundation including 0742818, 0742592, 1044978, 1110087, 1145172, 1145143, 1145248, 1639040, 1638957, 1638978, 1638970, 1726917, 1313010, 1313062, 1313158, 1313287, 0960243, 1836010, 1056465, \& 1255358 and by the Keck Foundation. The development of antenna-coupled detector technology was supported by the Jet Propulsion Laboratory Research and Technology Development Fund and NASA Grants 06-ARPA206-0040, 10-SAT10-0017, 12-SAT12-0031, 14-SAT14-0009, 16-SAT16-0002, \& 18-SAT18-0017. Focal plane development and testing were supported by the Gordon and Betty Moore Foundation at the California Institute of Technology. Readout electronics were supported by a Canada Foundation for Innovation grant to the University of British Columbia. The computations in this paper were run on the Cannon cluster supported by the FAS Science Division Research Computing Group at Harvard University. The analysis effort at Stanford University and the SLAC National Accelerator Laboratory was partially supported by the Department of Energy, Contract DE-AC02-76SF00515. We thank the staff of the U.S. Antarctic Program and in particular the South Pole Station without whose help this research would not have been possible. Tireless administrative support was provided by Amy Dierker, Kathy Deniston, Sheri Stoll, Irene Coyle, Donna Hernandez, and Dana Volponi.

\FloatBarrier

\bibliography{report} 
\bibliographystyle{spiebib} 

\end{document}